\documentclass[12pt]{article}
\renewcommand{\baselinestretch}{1.25}
\usepackage{amsthm,amsmath,latexsym,amssymb,amsfonts,amscd}
\usepackage{graphics,lscape,fancyhdr,array,stmaryrd,euscript,wrapfig}
\usepackage{empheq,slashed,skak}
\usepackage{verbatim,slashed}
\numberwithin{equation}{section}
\usepackage{hyperref,setspace}
\usepackage{tikz-cd}
\usepackage{mathrsfs}
\usepackage[numbers,sort&compress]{natbib}
\usepackage[nottoc]{tocbibind}

\newcommand{\pl}{\partial}

\newcommand{\be}{\begin{align}}
\newcommand{\ee}{\end{align}}

\newcommand{\fud}[2]{{}^{#1}{}_{#2}\,}

\newcommand{\p}[1]{(\ref{#1})}
\newcommand{\bea}{\begin{eqnarray}}
\newcommand{\eea}{\end{eqnarray}}

\newcommand{\besubeqs}{\begin{subequations}}
\newcommand{\esubeqs}{\end{subequations}}

\usepackage{tensor}
\usepackage{todonotes}

\renewcommand{\bar}[1]{\overline{#1}}

\newcommand{\aplus}[1]{\langle \boldsymbol{#1}\rangle }
\newcommand{\aminus}[1]{[\boldsymbol{#1}] }
\newcommand{\bb}[1]{\boldsymbol{#1}}

\usepackage[all,cmtip]{xy}

\begin{document}
\pagenumbering{gobble}
\graphicspath{ {./pics/} }
\begin{center}
{\Large\bfseries 
Cubic action for Spinning Black Holes\\ 

\vspace{10pt}
from massive higher-spin gauge symmetry
\vspace{0.4cm}
} \\

\vskip 0.04\textheight

Evgeny Skvortsov,\footnote{Research Associate of the Fund for Scientific Research -- FNRS, Belgium}\footnote{Also on leave from Lebedev Institute of Physics.}${}^{\symknight}$ and Mirian Tsulaia${}^{\symking}$

\vskip 0.04\textheight

{\em ${}^{\symknight}$ Service de Physique de l'Univers, Champs et Gravitation, \\ Universit\'e de Mons, 20 place du Parc, 7000 Mons, 
Belgium}\\

\vspace{5pt}
{\em$^{\symking}$ Okinawa Institute of Science and Technology, \\ 1919-1 Tancha, Onna-son, Okinawa 904-0495, Japan}

\vskip 0.02\textheight

{\bf Abstract }

\end{center}
\begin{quotation}
Scattering of two Kerr Black Holes emitting gravitational waves can be captured by an effective theory of a massive higher-spin field interacting with the gravitational field. While other compact objects should activate a multitude of non-minimal interactions it is the black holes that should be captured by the simplest minimal interaction. Implementing massive higher-spin symmetry via a string-inspired BRST approach we construct an action that reproduces the correct cubic amplitude of Arkani-Hamed--Huang--Huang. The same is achieved for the root-Kerr theory, i.e. for the minimal electromagnetic interaction of a massive higher-spin field.
\end{quotation}

\pagestyle{empty}
\newpage
\setcounter{page}{1}
\pagestyle{plain}
\renewcommand{\baselinestretch}{1.25}\normalsize
\newpage

\section{Introduction}
\pagenumbering{arabic}
\setcounter{page}{2}

Recent discovery of gravitational waves by LIGO/Virgo collaboration~\cite{LIGOScientific:2016aoc} has triggered a lot of interest in developing efficient techniques to model binary systems taking into account the gravitational radiation they emit. In addition to the canonical approaches based on solving Einstein equations one can employ the Effective Field Theory approach, where a compact object with angular momentum is represented as a massive particle with spin. The latter allows one to take advantage of numerous scattering amplitudes techniques, see e.g. \cite{Buonanno:2022pgc,Goldberger:2022ebt}.  
 
Massive higher-spin (non-elementary) particles do exist in nature in the form of hadrons, nuclei, etc. or can be used to model spinning objects within the effective field theory approach. As long as electromagnetic and gravitational fields are not strong enough to tear them apart they can be thought of as elementary higher-spin particles. It is quite amusing that even such macroscopical objects as black holes, neutron stars etc. can be modeled by massive higher-spin fields as long as the parameters of dynamics justify the point-like approximation.

At present, it seems plausible that the theory that describes the conservative dynamics of Kerr black holes is, in some sense, the simplest theory of a massive spin-$s$ field that couples it to gravity. Via the classical double copy construction \cite{Monteiro:2014cda,Arkani-Hamed:2019ymq,Guevara:2020xjx,Bern:2010ue} one can also think about the simplest theory of a spin-$s$ field interacting with photons/gluons, which bears the name root-Kerr. The main link between the black hole dynamics and field theory at the lowest nontrivial order is as follows. In \cite{Arkani-Hamed:2017jhn} an amplitude $s-s-h$ (AHH-amplitude) with the best high energy behavior that couples massive spin-$s$ field to gravity, $h=2$, or photons/gluons, $h=1$, was constructed. In the language of the massive spinor-helicity  \cite{Conde:2016vxs,Arkani-Hamed:2017jhn} it reads
\begin{align}
    \mathcal{A}(s,s,h^+)&= \tfrac{1}{m^{2s}}\mathcal{A}(0,0,h^+)\, \aplus{12}^{2s} & 
    \mathcal{A}(s,s,h^-)&= \tfrac{1}{m^{2s}}\mathcal{A}(0,0,h^-)\, \aminus{12}^{2s} 
\end{align}
where $\mathcal{A}(0,0,h)$ is the minimal coupling of the scalar field to the helicity-$h$ massless boson. This amplitude was shown to reproduce the dynamics of Kerr black holes \cite{Guevara:2018wpp,Chung:2018kqs,Guevara:2019fsj} at the lowest nontrivial order, which sets a challenge to construct a theory that starts with AHH-amplitude. 

While existing in reality, massive higher spin fields have so far been quite resistant to any theoretical attempts to introduce consistent interactions, the main problem being that almost any putative vertex activates unphysical degrees of freedom. In particular, the minimal interaction recipe $\pl \rightarrow D$ does not work already for electromagnetic interactions of $s=1$ field. 

There are several general approaches to the problem that are available at present. (i) Massive higher-spin gauge symmetry, i.e. massive fields can be described as theories with St\"uckelberg gauge symmetries that help to control the number of degrees of freedom. This approach has systematically been applied in \cite{Zinoviev:2001dt,Zinoviev:2006im,Zinoviev:2008ck,Zinoviev:2009hu,Zinoviev:2010cr,Buchbinder:2012iz} and the recent \cite{Cangemi:2022bew,Cangemi:2022abk}. (ii) Chiral approach \cite{Ochirov:2022nqz} where there is no danger to activate unphysical modes, but the parity is more difficult to control, see also \cite{Cangemi:2022bew,Cangemi:2022abk}. (iii) Light-cone approach \cite{Metsaev:2005ar,Metsaev:2007rn,Metsaev:2022yvb}; (iv) other, see e.g. \cite{Kuzenko:1994ju,
Buchbinder:2005ua,Buchbinder:2007ix,
Francia:2007ee,
Kaparulin:2012px,Sharapov:2023iwc, Buchbinder:2022lsi} for an incomplete list.

In the literature there are several different ways to represent massive higher-spin symmetry: (a) it can operate on a set of double-traceless fields $\Phi_{\mu_1 \ldots \mu_k}$, $k=0,\ldots,s$, \cite{Zinoviev:2001dt}; (b) these fields can be repackaged into a quartet of traceful tensors of ranks $s,s-1,s-2,s-3$ \cite{Pashnev:1989gm, Buchbinder:2008ss, Asano:2019smc,Lindwasser:2023zwo}; (c)  massive higher-spin gauge symmetry can also be imposed via Ward identities \cite{Cangemi:2022bew,Cangemi:2022abk}; (d) BRST approach
\cite{Hussain:1988uk, Pashnev:1997rm, Burdik:2000kj, Bekaert:2003uc, Francia:2010qp,Metsaev:2012uy}
takes its roots \cite{Witten:1985cc, Neveu:1985sh, Neveu:1986mv, Gross:1986ia} in string field theory. 

Zinoviev's and chiral approaches have recently been applied to the root-Kerr problem in \cite{Cangemi:2022bew,Cangemi:2022abk} with the main results being the quartic amplitude that passes all tests and an on-shell cubic action that gives AHH-amplitude. In this paper we implement massive higher-spin symmetry in a different way, via BRST, (d). Even though there are some $b$-$c$ ghosts in the system, there are no fields of opposite statistics present and BRST here is instrumental in dealing with auxiliary fields that simplify the formulation. The main result is a completely off-shell gauge-invariant action for Kerr and root-Kerr black holes together with the deformations of the gauge symmetry. 

The main advantage of the BRST approach is some sort of spin universality. The spin universality is the property of the black hole amplitudes that allows to extract a number of multipoles constrained from above by $s$ without having to take $s\rightarrow\infty$ limit,  \cite{Holstein:2008sw,Vaidya:2014kza, Aoude:2022trd, Cangemi:2022abk}. In other words, the Taylor coefficients of the multipole expansion do not depend on spin, but a spin-$s$ system can see only a finite number of them. In the BRST approach, to be precise, the set of consistent interactions for all spins turns out to be an associative algebra that is generated by a number of simple operators that couple triplets of low-spin fields. In this sense, the simplest interactions do no depend on spin as a parameter. Another useful feature is the simplicity of uplifting the on-shell results to off-shell ones. In fact, our starting point is a generating function of on-shell amplitudes from \cite{Chiodaroli:2021eug}.

The outline of the paper is as follows. In section \ref{sec:freeBRST} we review the BRST approach to free massive and massless higher spin fields. In section \ref{sec:cubic} we discuss the details of cubic interactions and the cubic action for a Kerr Black Hole is obtained in section \ref{sec:bhaction}. Conclusions and discussion can be found in section \ref{sec:conclusions}.

\section{BRST approach to massive and massless fields}
\label{sec:freeBRST}
The BRST approach reduces the problems of free fields and of interactions to $Q^2=0$ for a certain $Q$. The formalism simplifies a lot if instead of an irreducible spin-$s$ one considers a direct sum of massless or massive states with spins $s,s-2,...1/0$. 

\subsection{General formalism}

\paragraph{Massless fields.} We will need only massless fields with $s=1,2$, photon and graviton, but it is easier to treat all spins at once. A single spin-$s$ massless field can be described by a symmetric and traceless rank-$s$ tensor $\Phi_{\mu_1...\mu_s}(x)$, $\Phi\fud{\nu}{\nu\mu_3...\mu_s}=0$. On-shell it satisfies $\square \Phi_{\mu_1... \mu_s}=0$ and the physical states require a quotient by a gauge symmetry $\delta \Phi_{\mu_1...\mu_s}=\pl_{(\mu_1} \xi_{\mu_2...\mu_s)}$, with traceless gauge parameter $\xi_{\mu_1...\mu_{s-1}}$ being on-shell as well. If we drop the trace constraints we obviously get degrees of freedom of massless fields with spins $s,s-2,...,1/0$.

In order to hide the tensor indices it is often convenient to use an auxiliary
Fock space spanned by oscillators
\begin{equation}\label{B4-a-f}
[\alpha_{\mu}, \alpha_{\nu}^{ +} ] = - 
\eta_{\mu \nu}\,, \qquad \eta_{\mu \nu} = (+1,-1,-...,-1)\,,
\end{equation}
and define the Fock vacuum as
\begin{equation}
\alpha_\mu |0 \rangle_{\alpha}=0\,.
\end{equation}
A single field $\phi_{\mu_1 \mu_2... \mu_s}(x)$ or a collection of such fields with ranks from $0$ to $\infty$ is represented by a Fock-vector
\begin{equation}
| \phi \rangle = \frac{1}{s!} \phi_{\mu_1 \mu_2... \mu_s}(x)
\alpha^{\mu_1, +} \alpha^{\mu_2, +} ... \alpha^{\mu_s, +}|0 \rangle_{\alpha}\,.
\end{equation}
The d`Alambertian, divergence and symmetrized gradient operators are given by
\besubeqs
\begin{align}
 &\phi_{\mu_1 \mu_2... \mu_s}(x) && \leftrightarrow && \Box | \phi \rangle\,, \\
&\partial^\mu \, \phi_{\mu, \mu_2... \mu_s}(x) && \leftrightarrow && (\alpha \cdot p) \, |\phi \rangle\,, \\
&\partial_{(\nu} \phi_{\mu_1 \mu_2... \mu_s}(x) && \leftrightarrow && (\alpha^+ \cdot p) \, |\phi \rangle \,,
\end{align}
\esubeqs
with $p_\mu = \partial_\mu$; also we defined 
$\Box= p\cdot p$, 
and $A \cdot B \equiv A^\mu B_\mu$.
In order to construct a Lagrangian, which gives the mass-shell and transversality
conditions as a result of equations of motions we use the BRST method. We first
compute the algebra between the above mentioned operators, the only nonzero commutator being
\begin{equation}
[(\alpha \cdot p), (\alpha^+ \cdot p)  ] = - \Box\,.
\end{equation}
Then we introduce Grassmann -- odd nilpotent ghost variables $c_0, c$ and $c^+$ with ghost number 
$+1$ and
 $b_0, b^+$ and $b$, with ghost number $-1$. The only nonzero anticommutation relations between these
 variable are
\begin{equation} \label{B4-c-f}
\{ c^{ +}, b \} = \{ c, b^{+} \} 
= 
\{ c_{0} , b_{0} \} = 1 \,.
\end{equation}
Then one constructs a nilpotent BRST charge
 \begin{equation} \label{BRST-tr}
 Q = c_0 
 \Box
  - c ( \alpha^{+} \cdot p)
 + 
c^{+} (\alpha \cdot p)
  -  c^{+} c b_0, \qquad Q^2=0,
\end{equation}
which has the ghost number $+1$.
Now one can write a gauge invariant free Lagrangian
\begin{equation} \label{L-free}
{ \cal L}_{2} = - \int dc_0 
\langle \Phi | Q | \Phi
\rangle, \qquad\qquad \delta  |\Phi \rangle = Q |\Lambda \rangle\,,
\end{equation}
where integration over the Grassmann variable $c_0$ is defined as $\int dc_0 \, c_0=1$.

To perform a further analysis of the Lagrangian and of the field equations, 
one considers an extended Fock space with a total vacuum being
\begin{equation}
| 0 \rangle = | 0 \rangle_\alpha \otimes | 0 \rangle_{gh.} \qquad
c|0 \rangle_{gh.}=b|0 \rangle_{gh.}=b_0|0 \rangle_{gh.}=0\,.
\end{equation}
Therefore a field  $|\Phi \rangle $ 
and a parameter of gauge transformations $|\Lambda \rangle $
in the extended Fock space are expanded in terms of all
creation operators. Requiring the field 
to have the ghost number $0$
and the parameter of gauge transformations to have the ghost number
$-1$, we obtain
the expansion
in terms of the anticommuting operators
\begin{equation} \label{gh-exp-f}
    |\Phi \rangle = |\phi \rangle +
  c_0 b^{+}  |C \rangle +
  c^{+} b^{+}|D^{} \rangle, \qquad |\Lambda \rangle = b^{+} |\rho \rangle\,.
\end{equation}
Where the component fields $|\phi \rangle, |C \rangle,
|D  \rangle$ and parameters of gauge transformations
$|\rho \rangle$ 
are in turn expanded in terms 
of the oscillators
$\alpha^+_{\mu}$ only. Putting the expansion \p{gh-exp-f} into the  Lagrangian
\p{B4-c-f}, 
and eliminating anti-commuting variables via normal ordering and Grassmann integration,
we get the Lagrangian in terms of the component fields
\begin{align} \label{L-T-F}
{\cal L}_2&= - \langle \phi | \Box | \phi \rangle + \langle D |\Box | D\rangle - \langle C| C \rangle + 
 \\ \nonumber
& + \langle C| (\alpha^+ \cdot p) | D\rangle  +  \langle C| (\alpha \cdot p) | \phi \rangle 
 - \langle D | (\alpha \cdot p)| C\rangle  
-  \langle \phi| (\alpha^+ \cdot p)| C \rangle \,.
\end{align}
 The equations of motion that follow from this Lagrangian are
 \besubeqs
\begin{align} 
& \Box | \phi \rangle + (\alpha^+ \cdot p)  | C \rangle  =0 \,,\label{Ens-1} \\  
&  \Box | D\rangle  - (\alpha \cdot p) | C \rangle   =0\,, \label{Ens-2} \\  
& | C \rangle - (\alpha^+ \cdot p) | D \rangle - (\alpha \cdot p) | \phi \rangle =0\,.  \label{Ens-3}
\end{align}
\esubeqs
Similarly, using the equations \p{L-free} and \p{gh-exp-f} we get the gauge transformation
rules
 \begin{equation} 
 \delta |\phi \rangle = -(\alpha^+ \cdot p) | \rho \rangle,
 \qquad
 \delta | D \rangle = (\alpha \cdot p) | \rho \rangle, \qquad
  \delta | C \rangle  = \Box | \rho \rangle\,.
  \end{equation}
As it can be seen from the Lagrangian \p{L-free},
 fixing the number of the oscillators in
the field $|\phi \rangle$ to be equal to $s$, results into 
$| C \rangle$ and $| D \rangle$
having the number of the oscillators equal  to $s-1$ and
$s-2$ respectively. The  field $| C \rangle$ does not have 
a kinetic term, it is auxiliary. It can be either expressed in terms of the other two fields
via its own equation of motion  \p{Ens-3} and put back into the Lagrangian,
or can be gauged away, thus partially fixing the gauge as $\Box | \rho \rangle=0$.
The field  $| D \rangle$ is also auxiliary and it can be
gauged away.
As the result, one is left with the only physical field $|\phi \rangle$, which obeys the
mass-shell and transversality conditions. Let us note, that this gauge fixing procedure is not unique. For example, one could choose the light-cone gauge fixing straight away and reach the same conclusions.

\paragraph{Irreducible fields.}  One can construct
  a free Lagrangian for a single
 irreducible higher spin mode from the Lagrangian given in
  \p{L-free},
   by imposing an extra off-shell condition
\begin{equation} \label{ttcond}
 T|\Phi \rangle=0\,, \qquad T = - \tfrac{1}{2} 
  \alpha \cdot \alpha  - bc\,, \qquad [T,Q]=0\,.
\end{equation}
 This condition implies for the component fields
 \begin{equation}
 (\alpha \cdot \alpha)  |\phi \rangle -
 2|D \rangle = 0\,, \quad 
 (\alpha \cdot \alpha ) |D \rangle=0\,, \quad
 (\alpha \cdot \alpha)  |C \rangle=0\,.
 \end{equation}
  The first equation expresses the nonphysical field
  $|D \rangle$ in terms of the trace of the physical field
  $|\phi \rangle$. The second equation implies that 
  $|\phi \rangle$ is double-traceless. The third one  contains
  no new information,  because of the equation \p{Ens-3}.
  Now, expressing  $|C \rangle$ and 
  $|D \rangle$ in terms of $|\phi \rangle$ and putting these expressions
  back into the  Lagrangian, one gets the Fronsdal Lagrangian \cite{Fronsdal:1978rb},
  for a single higher-spin mode.
In a similar manner, 
acting with the operator $T$
 gauge transformation rule 
given in
\p{L-free}, one can see
that the parameter of the gauge transformations becomes traceless
$(\alpha \cdot \alpha ) |\rho \rangle=0$.

\paragraph{Massive fields.} In order to describe the massive triplet, one can use the method of dimensional reduction as a tool.
In particular, one considers a massless triplet in $d+1$ dimensions,
and decomposes
\begin{equation} \label{decomp}
\alpha^\pm_\mu \rightarrow (\alpha^\pm_{\hat \mu}\,, \xi^\pm),
\qquad
\Box \rightarrow \Box + m^2, \qquad \alpha^\pm \cdot p \rightarrow
\alpha^\pm \cdot p \pm m \xi^\pm\,,
\end{equation}
to obtain
\bea \label{BRST-m}
 Q_{\text{mass.}} &=& c_0 (
 \Box +m^2)
  - c (\alpha^{+} \cdot p + m \xi^+)
 + c^{+} (\alpha \cdot p - m \xi)
  -  c^{+} c b_0\,.
 \eea
All equations for the massive triplet, such as Lagrangian, gauge transformations
etc., can be obtained from the corresponding equations for the massless one,
by using the decomposition \p{decomp}, so we shall not present them here
(see for the details \cite{Pashnev:1997rm}--\cite{Bekaert:2003uc}).

  Before concluding this section let us make one  comment.
  As it was shown in \cite{Fotopoulos:2009iw}, after elimination
  of the field $| C \rangle$ via its own equations of motion,
  one can also represent the fields  $\phi^{[s]}$ and $D^{[s]}$
  as expansions in terms of double traceless fields
  $\Psi^{[s-2k]}$ where $0 \leq k \leq [\frac{s}{2}]$
  and of of their traces. 
  Substituting these expansions into the Lagrangian  \p{L-free}
  one obtains a sum of Fronsdal Lagrangians for the spins
  $s,s-2,..,1/0$. Again, formally doing this decomposition 
  for a massless triplet in
  $d+1$ dimensions and performing a dimensional reduction
  of the obtained sum of  Fronsdal Lagrangians, 
  one obtains a gauge invariant description of massive higher spin modes with spins 
  $s,s-2,..,1/0$ in
  $d$ dimensions.\footnote{It has been shown \cite{Pashnev:1989gm}, (see also \cite{Fotopoulos:2008ka},
  \cite{Bekaert:2003uc} for details
  and  for \cite{Buchbinder:2008ss} an alternative formulation), that one can obtain  Lagrangians for free irreducible massive higher-spin fields in
$d$ dimensions by a dimensional reduction of $d+1$ dimensional Fronsdal Lagrangians.} 

\subsection{Examples}

\paragraph{Massless $s=1$.} 
Since the case of a scalar is trivial\footnote{See
\cite{Buchbinder:2021qkt} for application of the BRST formalism
for description of various low spin systems,
such as $N=1$ Super Yang-Mills and SUGRA's.}, we start with a massless triplet with $s=1$.
The expansion \p{gh-exp-f} has the form
\begin{equation} \label{exp1-m0}
| \Phi \rangle = (\phi_\mu(x) \alpha^{\mu+} 
+  C(x) c_0 b^+)|0\rangle, \qquad
| \Lambda \rangle =  \rho(x) b^+|0\rangle\,.
\end{equation}
Inserting this expansion 
into the Lagrangian \p{L-T-F} and performing
the normal ordering, one obtains
the gauge invariant free Lagrangian 
\begin{equation} \label{Lm1-m0}
{\cal L}_2 =  \phi^\mu  \Box  \phi_\mu - C^2 - 2C \partial^\mu \phi_\mu \,,
\end{equation}
which is invariant under the transformations
\begin{equation} \label{gt-1-m0}
\delta \phi_\mu = -\partial_\mu \rho\,, \qquad
\delta C = \Box  \rho\,.
\end{equation}
The field equations obtained from the Lagrangian
\p{Lm1-m0}
are
\begin{equation} \label{fe-1-m0}
 \Box  \phi_\mu + 2\partial_\mu C=0,
\qquad
\partial^\mu \phi_\mu +C=0
\end{equation}
If one eliminates the auxilary field $C(x)$ 
via its own equation of motion, then one gets the Maxwell
Lagrangian. Alternatively, one can  eliminate the field $C(x)$ via the gauge transformations, thus obtaining a Lorentz gauge, or use the light-cone gauge fixing, by eliminating the component $\phi_+$ via the gauge transformations 
\p{fe-1-m0}.
The field equations \p{fe-1-m0} then imply,
 that the massless triplet for $s=1$
describes a massless vector field with transverse polarizations.

\paragraph{Massive $s=1$. } 
For the massive $s=1$ triplet (see the  discussion around the equations \p{decomp}-\p{BRST-m}) one has
\begin{equation} \label{exp1}
| \Phi \rangle = (\phi_\mu(x) \alpha^{\mu+} + \phi(x) \xi^+
+  C(x) c_0 b^+)|0\rangle\,, \qquad
| \Lambda \rangle =  \rho(x) b^+|0\rangle\,.
\end{equation}
After elimination  of the oscillators by normal ordering in \p{L-T-F},
one obtains
the gauge invariant free Lagrangian 
\begin{equation} \label{Lm1}
{\cal L}_2 = \phi^\mu ( \Box + m^2) \phi_\mu
- \phi ( \Box + m^2) \phi - C^2 -2C(\partial^\mu \phi_\mu + m\phi)\,,
\end{equation}
which is invariant under the transformations
\begin{equation} \label{gt-1}
\delta \phi_\mu = -\partial_\mu \rho\,, \qquad
\delta \phi = - m \rho\,, \qquad
\delta C = (\Box +m^2) \rho\,.
\end{equation}
The field equations obtained from the Lagrangian
\p{Lm1}
are
\besubeqs
\begin{align} \label{fe-1}
( \Box + m^2) \phi_\mu + \partial_\mu C= ( \Box + m^2) \phi + mC=0\,,
\\ \label{fe-2}
m \phi +\partial^\mu \phi_\mu +C=0\,.
\end{align}
\esubeqs
Similarly to the massless case, 
one can express the field $C(x)$ in terms of the other fields, 
and put it back into  \p{Lm1}, or
 gauge it away.
Next, one can 
 gauge away the field $\phi(x)$
(the procedure being consistent due to the equations of motion)
 thus using up the entire gauge freedom.
Then the transversality condition \p{fe-2} removes 
$\phi_0(x)$, whereas the first in the field equations
is the Klein--Gordon equation for the physical components of the vector field.

Alternatively,  one can again remove $C(x)$ first, but then instead of gauging away $\phi(x)$,
one uses
the gauge transformations \p{gt-1}, to remove $\phi_+(x)$. Then, the transversality condition removes $\phi_{-}(x)$, but the field
$\phi(x)$ remains intact
and plays the role of the
extra degree of freedom, which distinguishes a massive field from a massless one.
This gauge fixing procedure will be relevant to our discussion of cubic vertices
in the next Section.
 
\paragraph{Massless $s=2$.} 
For the massless $s=2$ triplet we have (see \cite{Pashnev:1997rm} for the massive one)
\begin{equation} \label{exp1-m0-2}
|\Phi \rangle = (\phi^{\mu \nu}(x) \alpha^{+}_\mu \alpha^{+}_\nu + c_0 b^+  C^\mu(x) \alpha^{+}_\mu 
+c^+ b^+ D(x)) |0 \rangle\,,  
\qquad
|\Lambda \rangle =  b^+ \rho^\mu(x) \alpha^{+}_\mu|0 \rangle\,.
\end{equation}
Similarly to the previous examples, we obtain from \p{L-T-F} the Lagrangian
\begin{equation} \label{lv2}
{\cal L}_2=  -\frac{1}{2}  \phi^{\mu \nu} \Box \phi_{\mu \nu}
+ D \Box D +  C^\mu C_\mu 
+ 2C^\mu \partial^\nu \phi_{\mu \nu} 
 + 2 D \partial^\mu C_\mu\,,
\end{equation}
which is invariant under gauge transformations
\begin{equation} \label{s2-1-1}
\delta \phi_{\mu \nu}  = -\partial_\mu \lambda_\nu - \partial_\nu \rho_\mu, \quad \delta C_\mu = \Box \rho_\mu, \quad \delta D = -\partial^\mu \rho_\mu\,,
\end{equation}
and  gives the equations of motion
\besubeqs
\begin{align} \label{s2-0}
&\Box \phi_{\mu \nu} + \partial_\mu C_\nu + \partial_\nu C_\mu=0\,, \\ \label{s2-2}
&\Box D + \partial^\mu C_\mu=0\,, \\
&C - \partial_\mu D+
\partial^\nu \phi_{\mu \nu}  = 0 \,.
\label{s2-3}
\end{align}
\esubeqs
As in the case of the previous examples on can gauge away the fields
$C_\mu(x)$ and $D(x)$ (or use the light-cone gauge fixing) to obtain
the Klein-Gordon equations for massless field $\phi_{\mu \nu}(x)$.
This field describes simultaneously spins $2$ and $0$, due to the lack of transversality
condition. This can be also seen by introducing new fields $\Psi_{\mu \nu}(x)$
and $\Psi(x)$ (see the discussion at the end of the previous Section)
\begin{equation}
\phi_{\mu \nu} = \Psi_{\mu \nu} + \frac{\eta_{\mu \nu}}{d-2} \Psi\,, \qquad\qquad
\phi^\mu{}_\mu + 2D = \Psi\,.
\end{equation}
Substituting these expressions into \p{lv2}, one can see that it splits
into a sum
of two Fronsdal Lagrangians, one for the field $\Psi_{\mu \nu}(x)$ and one for the field $\Psi(x)$.  
If one adds an extra off-shell condition \p{ttcond}, which in this case reads
$\Psi(x)=0$, then one obtains the Fronsdal Lagrangian for spin-$2$, thus eliminating
the lower spin component.

\section{Cubic Action}
\label{sec:cubic}
On the way towards the cubic action of a Kerr Black Hole we discuss a well-developed formalism, see \cite{Metsaev:2012uy}, to construct cubic interactions within the BRST approach. The main advantage is that the vertices can be represented as polynomials in a few atomic structures. After the preliminaries, we engineer the right cubic action.

\subsection{Generators of Cubic vertices}
In order to construct cubic interactions
of two fields with spin-$s$ and mass-$m$ and
a massless field, 
we take three copies of the auxiliary Fock space spanned by  the oscillators 
\besubeqs
\begin{equation}\label{B4-a}
[\alpha_{\mu}^{(i)}, \alpha_{\nu}^{(j), +} ] = - \delta^{ij}
\eta_{\mu \nu}\,,
\qquad
[\xi^{(i)}, \xi^{(j), +} ] =  \delta^{ij}\,,
\end{equation}
\begin{equation} \label{B4-c}
\{ c^{(i), +}, b^{(j)} \} = \{ c^{(i)}, b^{(j),+} \} 
= 
\{ c_{0}^{(i)} , b_{0}^{(j)} \} =
 \delta^{ij}\,,
\end{equation}
\esubeqs
$$
 i,j=1,2,3\,.
$$
We put the massive fields into the Fock spaces with $i=1,2$
and massless fields into the Fock space with $i=3$.  
The fields $|\Phi^{(i)} \rangle$ and the parameters
of gauge transformations 
$|\Lambda^{(i)} \rangle$
have the same expression
as \p{gh-exp-f}
\begin{equation} \label{gh-exp}
    |\Phi^{(i)} \rangle = |\phi^{(i)} \rangle +
  c^{(i)}_0 b^{(i),+}  |C^{(i)} \rangle +
  c^{(i),+} b^{(i),+}|D^{(i)} \rangle\,, \qquad |\Lambda^{(i)} \rangle = b^{(i),+} |\rho^{(i)} \rangle\,,
\end{equation}
with no summation over $i$.
The component fields $|\phi^{(i)} \rangle, |C^{(i)} \rangle,
|D^{(i)}  \rangle$ and parameters of gauge transformations
$|\rho^{(i)} \rangle$ 
are in turn expanded in terms 
of the oscillators
$\alpha^{(i),+}$ and $\xi^{(1,2),+}$ (for massive fields).
The fields and parameters of gauge transformations
can also have indices, which correspond to an internal symmetry group. We shall not write them explicitly,
and assume their presence when the sum of the spins in a cubic vertex is odd.

Let us consider  Lagrangian with cubic interactions included
\bea \label{LNSINT}
{ \cal L}_{3} &=& -\sum_{i=1}^3 \int dc_0^{(i)} 
\langle \Phi^{(i)} |Q^{(i)}| \Phi^{(i)}
\rangle + \\ \nonumber 
&+&g \left ( \int dc_0^{(1)} dc_0^{(2)} dc_0^{(3)} \langle \Phi^{(1)}| \langle \Phi^{(2)}| \langle \Phi^{(3)}| |V \rangle + h.c.
\right )\,,
\eea
where $g $ is a coupling constant.
The nilpotent BRST charges are  the same as in
\p{BRST-m} for massive fields,
and \p{BRST-tr}
for massless fields, i.e.,
\besubeqs
\begin{align}
\label{BRST-m-2}
 Q^{(1,2)} &= c_0^{(1,2)} (
 \Box^{(1,2)} +m^2)
  - c^{(1,2)} (\alpha^{(1,2),+} \cdot p^{(1,2)} + m \xi^{(1,2),+})
 + \\ \nonumber
 &+c^{(1,2),+} (\alpha^{(1,2)} \cdot p^{(1,2)} - m \xi^{(1,2)})
  -  c^{(1,2),+} c^{(1,2)} b^{(1,2)}_0\,,
\\ \label{BRST-m-0}
 Q^{(3)} &= c_0^{(3)} 
 \Box^{(3)} 
  - c^{(3)} \alpha^{(3),+} \cdot p^{(3)} 
 + c^{(3),+} \alpha^{(3),} \cdot p^{(3)} 
  -  c^{(3),+} c^{(3),} b^{(3)}_0\,. 
 \end{align}
\esubeqs
 The cubic vertex which is present in the Lagrangian
 \p{LNSINT}
 has the form
\begin{equation} \label{V3Bosons}
|V \rangle = V(p_\mu^{(i)}, \alpha_{\mu}^{(i),+},
\xi^{(i),+}
c^{(i), +}, b^{(i), +}, b^{(i)}_{0} ) \,\, c_0^{(1)} c_0^{(2)} c_0^{(3)} \,, 
| 0^{(1)} \rangle \otimes
|0^{(2)} \rangle \otimes | 0^{(3)} \rangle\,,
\end{equation}
where the function $V$  is restricted to have the ghost number zero, to be a Lorentz invariant and 
to be a solution of
the BRST invariance condition
\begin{equation} \label{NSBRSTV}
(Q^{(1)} + Q^{(2)}+ Q^{(3)})  | V  \rangle=0\,.
\end{equation}
This condition is obtained from
 the requirement of  invariance of the Lagrangian \p{LNSINT} up to the terms linear in $g$
 under the non-linear gauge transformations
\bea \label{GTNSINT}
\delta | \Phi^{(i)} \rangle & = &  Q^{(i)} |\Lambda^{(i)} \rangle
+  \\ \nonumber 
&+& g \int dc_0^{(i+1)} dc_0^{(i+2)} 
\left ( (\langle \Phi^{(i+1)}| \langle \Lambda^{(i+2)}|
+ \langle \Lambda^{(i+1)}| \langle \Phi^{(i+2)}| ) | V \rangle \right )\,,
\eea
where $i+1$, $i+2$ are understood modulo $3$. 
The  condition
\p{NSBRSTV}
guarantees also that the group structure of the gauge transformations is preserved up to the first order in $g$. Using momentum conservation
\begin{equation}  \label{mcon}
p_\mu^{(1)} + p_\mu^{(2)} + p_\mu^{(3)}=0\,,
\end{equation}
and the commutation relations \p{B4-a}-- \p{B4-c},
one can show,   
that the following atomic expressions are BRST invariant for any values of the spins entering the cubic vertex \cite{Metsaev:2012uy}
\besubeqs\label{atoms}
\begin{align}  \label{sbv-1}
{\cal K}^{(1)} &= 
-(p^{(2)}- p^{(3)}) \cdot \alpha^{(1),+} +
 m \xi^{(1),+} +
(b_0^{(2)}-b_0^{(3)}) \, c^{(1),+}\,,
\\  \label{sbv-2}
{\cal K}^{(2)} &= 
-(p^{(3)}- p^{(1)}) \cdot \alpha^{(2),+} 
- m \xi^{(2),+} +
(b_0^{(3)}-b_0^{(1)}) \, c^{(2),+}\,,
\\  \label{sbv-3}
{\cal K}^{(3)} &= 
-(p^{(1)}- p^{(2)}) \cdot \alpha^{(3),+} 
 +
(b_0^{(1)}-b_0^{(2)}) \, c^{(3),+}\,,
\\  \label{sbv-4}
{\cal Q} &=- \alpha^{(1),+} \cdot \alpha^{(2),+} +
\tfrac{\xi^{(1),+}}{2m} {\cal K}^{(2)}-
\tfrac{\xi^{(2),+}}{2m} {\cal K}^{(1)} + \xi^{(1),+}\xi^{(2),+}
- \tfrac{1}{2} b^{(1),+} c^{(2),+} - \tfrac{1}{2} b^{(2),+} c^{(1),+}
\end{align}
and
\begin{equation}  \label{sbv-5}
{\cal Z}= {\cal Q}^{(1,2)}{\cal K}^{(3)} +
{\cal Q}^{(2,3)} {\cal K}^{(1)} +
{\cal Q}^{(3,1)} {\cal K}^{(2)}\,,
\end{equation}
\esubeqs
where
\begin{equation}  \label{sbv-6}
{\cal Q}^{(12)} =- \alpha^{(1),+} \cdot \alpha^{(2),+}
+ \xi^{(1),+}\xi^{(2),+}
- \frac{1}{2} b^{(1),+} c^{(2),+} - \frac{1}{2} b^{(2),+} c^{(1),+}\,,
\end{equation}
\begin{equation}  \label{sbv-7}
{\cal Q}^{(i,i+1)} = - \alpha^{(i),+} \cdot \alpha^{(i+1),+}
- \frac{1}{2} b^{(i),+} c^{(i+1),+} - \frac{1}{2} b^{(i+1),+} c^{(i),+}\,,
\quad i=2,3\,.
\end{equation}
One can check, that the vertices \p{sbv-1}--\p{sbv-5}
are  BRST non-trivial solutions of the equation
\p{NSBRSTV}, and therefore they can not be obtained from the free Lagrangian
by field redefinition.
An obvious consequence of the BRST invariance of these atomic vertices is that any function thereof is a valid cubic vertex, unless some extra selection 
criteria are imposed.

\paragraph{Relation to the light-cone gauge vertices.} Before concluding the discussion
on cubic interactions, let us note, that  
the cubic vertices considered above are actually
covariantizations of the cubic vertices
given in the light-cone gauge 
in \cite{Metsaev:2005ar}. 
In order to see this
let us recall, that in the light-cone formalism 
one does not have the fields $|C \rangle$ and $|D \rangle$ (which are artifacts of the gauge invariant formulation),
 whereas the physical fields
 $| \phi \rangle$ have the form 
\begin{equation} \label{expL-C}
| \phi \rangle = \sum_{k=0}^s \frac{1}{(s-k)! k!}
\phi_{I_1  ..., I_{s-k}}(x)\alpha^{I_1+}...
\alpha^{I_{s-k}+}
(\xi^{+})^k|0\rangle\,,
\end{equation}
The light-cone index is $I=1,..,d-2$,   both for
massless and massive fields, 
whereas the oscillator $\xi^+$
plays the role of an ``extra" degree of freedom,
which is present in the massive representations.
Such representation for massive fields allows  one to  couple
them to massless ones  in  cubic vertices by  contractions of the indices $I_{k}$ between the fields and the corresponding momenta.

 Going back to the gauge-invariant description, one promotes
the  index
$I$ to the index $\mu=0,..,d-1$ at the expense of introducing 
 of the auxiliary fields
$|C \rangle$ and  $|D \rangle$.
Comparing explicit expressions
of the cubic vertices \p{sbv-1}--\p{sbv-5} is
with the ones found in the light-cone gauge
\cite{Metsaev:2005ar}, 
can immediately establish the direct correspondence between them
\cite{Metsaev:2012uy}.

The cubic vertices given above, correspond to interactions
of two massive  fields with equal masses with a massless
field, where the spins of all three fields can be arbitrary.
In the further discussion we shall restrict our attention
to massless triplets with the higher spin equal to $2$ and $1$.
To this end, we take the form of the fields in the third Fock space
as in \p{exp1-m0-2} and 
in \p{exp1-m0}
respectively. Also in the case of interactions
of massive fields  with the massless spin-$2$ triplet, we put identical
tensor fields ($\phi_{\mu_1,.., \mu_k}(x)$,
$C_{\mu_1,.., \mu_l}(x)$, $D_{\mu_1,.., \mu_m}(x)$)
in the first and the second Fock spaces.
Similarly, these fields will get an internal index
for the case of interaction with the massless spin-$1$ triplet.

\subsection{Simple three-point amplitudes}
Let us compute three-point amplitudes for two massive
fields with spin-$s$ and mass-$m$ with one  massless field with spin-$2$ or spin-$1$,  by using
the cubic vertices given here-above. We shall confine our attention to the case when the dimension
of the space-time is equal to four and use the spinor--helicity formalism (see, e.g., \cite{Conde:2016vxs,Arkani-Hamed:2017jhn} for the conventions and useful identities and Appendix \ref{app:A}).

The asymptotic ``in" and ``out" states are described by  physical
fields $|\phi^{(i)} \rangle$ 
with the total number of
$\alpha_\mu^{(i),+}$ and $\xi^{(i),+}$ for  $i=1,2$
is equal to $s$. As we discussed in the Section \ref{sec:freeBRST},
we can decompose this field into irreducible modes of spins
$s, s-2, \ldots 1/0$.
For the massless triplet with $i=3$ and $s=2$, we can similarly
decompose the physical field into the components with spins $2$ and $0$. The massless triplet with $s=1$ is already irreducible. First, let us note, that although the cubic vertices
\p{sbv-1} -- \p{sbv-5}
and massive triplets contain the  $\xi^{(1,2),+}$, one can 
use only dependence on $\alpha_\mu^{(1,2,3),+}$ when computing the three point amplitudes. Indeed, this is the only oscillator that can be contracted with a rank-$s$ polarization tensor. 

Let us consider some lower spin examples, which will explain the meaning of the atomic cubic structures \eqref{atoms}.  

\paragraph{The case $\boldsymbol{0-0-1}$.}
For this simplest case
 the fields are
\begin{equation}
\langle \phi^{(1)} | = \langle 0| \phi^{(1)}\,, \qquad
\langle \phi^{(2)} | = \langle 0| \phi^{(2)}\,, \qquad
\langle \phi^{(3)} |  = \langle 0 |  \alpha_\mu^{(3)} 
 \epsilon_\mu^-\,,
\end{equation}
where we have taken the massless vector field with the negative helicity.
Apparently, the relevant cubic  vertex has the form
\begin{equation} \label{c-001}
V= - \tfrac{i}{\sqrt 2}  \,{\cal K}^{(3)}\,.
\end{equation}
 Therefore, the three-point amplitude between two massive scalars and a massless vector
 is
\begin{equation}
A_{\phi \phi A^{-}}=  \tfrac{1}{\sqrt 2}\epsilon^- \cdot (p^{(1)}-p^{(2)}) =
{\sqrt 2} \epsilon^- \cdot p^{(1)} = mx^{-1}\,,
\end{equation}
where we have used the transversality of the vector and introduced the $x$-factor according
to \cite{Arkani-Hamed:2017jhn}. For instance, this is an amplitude between the complex scalar field and a photon induced by the minimal current interaction. 

\paragraph{The case $\boldsymbol{1-1-1}$.}
For two massive and one massless vector fields we have
\begin{equation}
\langle \phi^{(1)} | = \langle 0| \alpha_\mu^{(1)} 
 \epsilon_\mu^{(1)}
\,, \qquad
\langle \phi^{(2)} | = \langle 0| \alpha_\mu^{(2)} 
 \epsilon_\mu^{(2)}\,, 
\qquad
\langle \phi^{(3)} |  = \langle 0 |  \alpha_\mu^{(3)} 
 \epsilon_\mu^{(3),-}\,.
\end{equation}
The relevant cubic vertex is
\begin{equation}
V= - \tfrac{i}{\sqrt 2} {\cal Z}\,,
\end{equation}
which gives the three-point amplitude as
\begin{equation}
A_{WWA} =i {\sqrt 2} 
\left ( (\epsilon^{(1)} \cdot \epsilon^{(2)})
(\epsilon^{(3)} \cdot p^{(2)})
+
(\epsilon^{(2)} \cdot \epsilon^{(3)})
(\epsilon^{(1)} \cdot p^{(3)})
+
(\epsilon^{(3)} \cdot \epsilon^{(1)})
(\epsilon^{(2)} \cdot p^{(1)}) \right )\,.
\end{equation}
Rewriting this amplitude in the spinor-helicity formalism by using
the relations from Appendix \ref{app:A}, we get
\begin{equation}
A_{WWA^{-}} = \frac{1}{mx} \aminus{12}^2 \,.
\end{equation}
For instance, this is an amplitude between the W-boson and a photon.

\paragraph{The case $\boldsymbol{1-1-2}$.}

 For the case of two massive vector fields and a graviton
we have 
\begin{equation}
\langle \phi^{(1)} | = \langle 0| \alpha_\mu^{(1)} 
 \epsilon_\mu^{(1)}
\,, \quad
\langle \phi^{(2)} | = \langle 0| \alpha_\mu^{(2)} 
 \epsilon_\mu^{(2)}\,, 
\quad
\langle \phi^{(3)} |  =\frac{1}{2} \, \langle 0 |  \alpha_\mu^{(3)} \alpha_\nu^{(3)} \,
 \epsilon_\mu^{(3),-} \, \epsilon_\nu^{(3),-}\,.
\end{equation}
The relevant vertex has the form 
\begin{equation}
V = - \tfrac{i}{ 2} {\cal K}^{(3)} {\cal Z}\,,
\end{equation}
which gives
\begin{equation}
A_{1,1,2} = i A_{WWA} A_{\phi \phi A} \,.
\end{equation}
Using the spinor-helicity formalism,
one obtains
\begin{equation}
A(\bb{1}, \bb{ 1},2^{-}) = i  \frac{\aminus{12}^{2}}{x^2}\,.
\end{equation}
For instance, this is the usual minimal interaction via the stress-tensor between a massive vector and a graviton. 

\paragraph{The case $\boldsymbol{2-2-2}$. } For the case of two massive spin-$2$ fields and one graviton, we have $\langle \phi^{(3)} |$  as before and 
\begin{equation}
\langle \phi^{(1,2)} |  =\frac{1}{2} \, \langle 0 |  \alpha_\mu^{(1,2)} \alpha_\nu^{(1,2)} \,
 \epsilon_\mu^{(1,2)} \, \epsilon_\nu^{(1,2)}\,.
\end{equation}
The relevant
 vertex has the form
\begin{equation} \label{gr-v}
V = - \tfrac{i}{ 2} {\cal Z}^2\,.
\end{equation}
 Let us note that there are some other possible
combinations of the atomic vertices \p{sbv-1}-\p{sbv-5},
which have the powers of the oscillators
equal to $(2,2,2)$.
The choice \p{gr-v} is singled out by the requirement
that the three-point amplitude
has the ``minimal form" given by the AHH-amplitude
\begin{equation}
A(\bb{2},\bb{ 2},2^{-}) = i \frac{m^2}{x^2} \frac{\aminus{12}^{4}}{m^{4}}\,,
\end{equation}
when written in terms of spinor-helicity variables.

\paragraph{The case $\bb{3-3-2}$.} The fields in this case are  $\langle \phi^{(3)} |$  as before  and
\begin{equation}
\langle \phi^{(1,2)} |  =\frac{1}{3!} \, \langle 0 | 
\alpha_\mu^{(1,2)} \alpha_\nu^{(1,2)} \alpha_\rho^{(1,2)}\,
 \epsilon_\mu^{(1,2)} \, \epsilon_\nu^{(1,2)} \epsilon_\nu^{(1,2)}\,.
\end{equation}
Again, in this case there are several possibilities for the cubic interaction vertex, each of them giving a valid gauge-invariant expression. 
 Therefore, we can
take  all combinations
of the vertices \p{sbv-1}-\p{sbv-5} 
with powers  of oscillators being $(3,3,2)$,
and add them up with arbitrary coefficients.
We choose the coefficients 
from the requirement
that the three point amplitude corresponds to
the AHH-amplitude. It turns out that such vertex has the form
\begin{equation}
V = -\tfrac{i}{2} \left (
{\cal Z}^2 {\cal Q} - {\cal Z} {\cal Q}^2 {\cal K}^{(3)}
+ {\cal Z}^2 {\cal K}^{(1)} {\cal K}^{(2)}\frac{1}{2m^2}
\right )
\,.
\end{equation}

\subsection{Action for a Kerr Black Hole}
\label{sec:bhaction}
One can generalize the lower spin examples, considered above to an arbitrary spin. While the relation between the atomic cubic vertices \eqref{atoms} and amplitudes has hopefully been made clear by the examples above, the main advantage of the formalism is that any polynomial in \eqref{atoms} leads automatically to an off-shell gauge-invariant action. This is thanks to the BRST formulation that relates vertices and gauge transformations via the auxiliary ghost variables and thanks to the triplet formulation that makes this relation simple. 

The general three point amplitude 
of two massive higher-spin fields with the graviton,
which gives, the AHH-amplitude, was obtained in
\cite{Chiodaroli:2021eug} 
\begin{equation} \label{amp-gene}
\sum_s A (\bb{s},\bb{s},2) = i  A_{\phi \phi A} A_{\phi \phi A}
+
iA_{WWA}
\left ( A_{\phi \phi A}+  \frac{A_{WWA} -
(\epsilon_1 \cdot \epsilon_2)^2A_{\phi \phi A}
}{ (1 + \epsilon_1 \cdot \epsilon_2)^2  + \frac{2}{m^2} (\epsilon_1 \cdot p_2) (\epsilon_2 \cdot p_1)}
\right)\,.
\end{equation}
Indeed, 
using the spinor identities, one gets
the following expression for the (chiral) three-point amplitude
\begin{equation} \label{amp-gene-s}
A(\bb{s},\bb{s},2^{-}) = i \frac{m^2}{x^2} \frac{\aminus{12}^{2s}}{m^{2s}}\,.
\end{equation}
With the help of the dictionary between the three-point amplitudes for lower spins
and cubic vertices \p{sbv-1}-- \p{sbv-5}, 
\begin{equation}\notag
A_{\phi \phi A} = - \frac{i}{\sqrt 2}{\cal K}^{(3)}, \quad
A_{WWA} =-  \frac{i}{\sqrt 2} {\cal Z}, \quad \epsilon_1 \cdot \epsilon_2 =
- {\cal Q}, \quad \epsilon_1 \cdot p_2 = -\frac{1}{2}
{\cal K}^{(1)}, 
\quad \epsilon_2 \cdot p_1 =  \frac{1}{2} 
{\cal K}^{(2)}\,,
\end{equation}
established above, one can immediately write the cubic Lagrangian vertex $V$ as 
\begin{equation} \label{gene}
V= \sum_s V(s,s,2)= - \tfrac{i}{2} ({\cal K}^{(3)})^2
- \tfrac{i}{2} 
\left ( {\cal Z}{\cal K}^{(3)}+  \frac{{\cal Z}^2 -
{\cal Q}^2 {\cal K}^{(3)} {\cal Z}
}{ (1 - {\cal Q})^2  - \frac{1}{2m^2} 
{\cal K}^{(1)} {\cal K}^{(2)}
}
\right)\,,
\end{equation}
which gives the amplitude \p{amp-gene} (or, equivalently, the amplitude \p{amp-gene-s}).
Similarly, the cubic vertex, 
which corresponds to the AHH three-point amplitude
for two massive field with spin-$s$, mass-$m$
and a massless vector field reads 
\begin{equation} \label{gene-1}
V=\sum_s V(s,s,1)= 
- \tfrac{i}{\sqrt 2}  {\cal K}^{(3)}
- \tfrac{i}{\sqrt 2} 
\frac{{\cal Z} -
{\cal Q}^2 {\cal K}^{(3)}
}{ (1 - {\cal Q})^2  - \frac{1}{2m^2} 
{\cal K}^{(1)} {\cal K}^{(2)}
}\,.
\end{equation}
The cubic vertices for Kerr \eqref{gene} and root-Kerr \eqref{gene-1} theories are the main results of the paper. Within the BRST formalism they also encode the corrections to the free gauge transformations via \eqref{GTNSINT}.
Let us note that using the light-cone versions
of the vertices \p{sbv-1}--\p{sbv-5} 
given in \cite{Metsaev:2005ar}, one can write
the same Lagrangians in the light-cone gauge.
The free part of the Lagrangian will contain
only kinetic terms for the physical fields
and the cubic interactions being 
\p{gene} and \p{gene-1}. Let us also note that our result is based on the generating function of the AHH-amplitudes found in \cite{Chiodaroli:2021eug}, where the individual spin-$s$ amplitudes were summed with coefficient $1$. In view of the coherent state formalism \cite{Aoude:2021oqj} it would also make sense to sum the amplitudes with weight $1/(2s)!$ to get the diagonal part of the spin-coherent amplitude. 

\section{Conclusions and Discussion}
\label{sec:conclusions}
In this paper we found an action for Kerr Black Holes up to the cubic order. The cubic action is rather simple in terms of a generating function and can almost be read off from the AHH-amplitude, which, on one hand, makes a lot of sense since an amplitude is the only physical information encoded in a vertex, on the other hand, the simplicity of the off-shell uplift is striking. As a result, we have at our disposal a cubic Lagrangian $\mathcal{L}_3$ together with the corrections $\delta_1$ to the gauge transformations. 

At the conceptual level the problems of massive and massless higher-spin fields seem to be rather different from each other. Interactions of massless higher-spins are severely constrained with a handful of theories making it to be perturbatively local field theories, see \cite{Bekaert:2022poo}. The massless higher-spin multiplets are usually infinite. On the contrary, massive higher-spin fields can interact with electromagnetism and gravity individually, they can also exhibit self-interactions without having to invoke other fields.  

Nevertheless, the study of massless higher-spin fields provides some insights into the problem of massive ones. For example, the higher energy limit of the cubic amplitudes of massive fields is governed by those of massless. On a different note, via the Stuckelberg approach a massive spin-$s$ field can be realized as collections of massless fields with spins from $0$ to $s$, which allows one to employ the usual massless higher-spin gauge symmetry to control the number of degrees of freedom, which is what has been done in the present paper, see also \cite{Cangemi:2022bew,Cangemi:2022abk}. 

The on-shell Compton amplitude that passes a number of nontrivial tests has recently been found in \cite{Cangemi:2022bew,Cangemi:2022abk}. However, extending this result to the action level (i.e. off-shell) is a challenge. We hope that the approach advocated in the present paper can simplify the problem and provide an off-shell uplift of the on-shell results of \cite{Cangemi:2022bew,Cangemi:2022abk}. 

Another feature of the BRST formalism, which may turn out to be useful for applications, is the reducibility of the multiplet: a single field describes particles with spins $s,s-2,\ldots 1/0$. Therefore, one can implement spin-changing interactions directly within the multiplet. However, mass-changing interactions would require a family of multiplets, see the recent \cite{Bern:2023ity,Aoude:2023fdm,Jones:2023ugm} for the first steps towards taking such interactions into account. The atomic building blocks of the cubic vertices can also be employed to construct spin- and mass-changing interactions.

\section*{Acknowledgments}
\label{sec:Aknowledgements}
We would like to thank Lucile Cangemi, Marco Chiodaroli, Henrik Johansson,
Yasha Neiman, Alex Ochirov, Paolo Pichini, Dmitri Sorokin for useful discussions and comments. The work of E.S. was partially supported by the European Research Council (ERC) under the European Union’s Horizon 2020 research and innovation programme (grant agreement No 101002551). The work of M.T. was supported by the Quantum Gravity Unit of the Okinawa Institute of Science and Technology Graduate University (OIST). 

\appendix
\section{(Massive) spinor-helicity} \label{app:A}
We mainly follow the conventions of \cite{Chiodaroli:2021eug, Chung:2018kqs}. The massless spinors are   $\lambda_\alpha$ and
$\bar \lambda_{\dot \alpha}$, where
$\alpha=1,2$ and $\dot \alpha=1,2$ are indices 
of the four dimensional Lorentz group
$SL(2,C)$.
We use a notation
\begin{equation}
| \lambda \rangle \leftrightarrow \lambda_\alpha, \quad
| \lambda ] \leftrightarrow \bar \lambda^{\dot \alpha}, \quad
\langle \lambda | \leftrightarrow \lambda^\alpha, \quad
[ \lambda | \leftrightarrow \bar \lambda_{\dot \alpha}
\end{equation}
Summations  over Lorentz indices are defined as
\begin{equation}
\langle i j \rangle = \lambda^{(i),\alpha}
\lambda_{\alpha}^{(j)}, \quad
[ i j] = \bar \lambda_{\dot \alpha}^{(i)} 
\bar \lambda^{ (j), \dot \alpha} \quad
\langle i \sigma^\mu j ] = \lambda^{(i), \alpha} \sigma^\mu_{\alpha \dot \alpha} 
\bar \lambda^{(j), \dot \alpha}
\end{equation}
where $i,j$ are numbers of the particles.

The massive spinors are $\lambda_\alpha^a$ and
$\bar \lambda_{\dot \alpha}^a$. The index 
$a=1,2$ corresponds to the little group $SU(2)$.
For the massive spinors we have
\begin{equation}
| \lambda_a \rangle_\alpha \, \langle^a \lambda |^\beta = m \delta_\alpha^\beta, \quad
| \lambda^a ]^\alpha \, [ \lambda_a |_\beta = m \delta^\alpha_\beta,
\quad [\lambda^a \lambda^b]= - \langle \lambda^a \lambda^b \rangle =
m \varepsilon^{ab}\,.
\end{equation}
Spinorial indices are raised and lowered as
$\lambda^a = \varepsilon^{ab}\lambda_b$ and
$\lambda_a = \varepsilon_{ab}\lambda^b$ with $\varepsilon^{12}=\varepsilon_{21}=1$, 
the same is for $\alpha$ and $\dot \alpha$
indices. The convention for the summation over Lorentz indices are
the same as for the massless ones.

As before, the labels $"1"$ and $"2"$ are for massive fields with the same mass $m$ and
the label $"3"$ is for the massless spin $2$ or spin $1$
field.
In a three-point vertex the momentum conservation law implies
\begin{equation}
p^{(1)}_{\alpha \dot \alpha} + p^{(2)}_{\alpha \dot \alpha} +
p^{(2)}_{\alpha \dot \alpha} = 
\lambda_\alpha^{(1), a} \bar \lambda^{(1)}_{\dot \alpha, a} \,
+
\lambda_\alpha^{(2), a} \bar \lambda^{(2)}_{\dot \alpha, a}\, +
\lambda_\alpha^{(3)} \bar \lambda^{(3)}_{\dot \alpha} =
0\,.
\end{equation}
It is helpful to introduce auxiliary
commuting spinors $z_a$ and consider
\begin{equation}
{  \lambda}_\alpha = \lambda_\alpha^a z_a, \qquad
{ \bar \lambda_{\dot \alpha}} = \bar \lambda_{\dot \alpha}^a z_a\,.
\end{equation}
The massive spinors satisfy the Dirac equation
\begin{equation}
p_{\alpha \dot \alpha } \bar \lambda^{\dot \alpha}
= m \lambda_\alpha, \quad
p_{\alpha \dot \alpha }  \lambda^{ \alpha}
= -
m \bar \lambda_{\dot \alpha}\,.
\end{equation}
Like massless spinors,
they also satisfy Schouten identity
\begin{equation}
\langle \lambda_1 \lambda_2 \rangle 
\langle \lambda_3 | +
\langle \lambda_3 \lambda_1 \rangle 
\langle \lambda_2 | +
\langle \lambda_2 \lambda_3 \rangle 
\langle \lambda_1 | =0
\end{equation}
as well as $\langle \lambda_1 \lambda_1 \rangle=
[ \lambda_1 \lambda_1 ]=0 $.

Vector and spinor indices are connected as 
$
p_{\alpha \dot \alpha } = p_\mu (\sigma^\mu)_{\alpha \dot \alpha} 
$
with
\begin{equation}
tr (\sigma^\mu \sigma^\nu) = 2 \eta^{\mu \nu}, \qquad
(\sigma^\mu)_{\alpha \dot \alpha} \,
(\sigma_\mu)^{\beta \dot \beta} = 2 
\delta_{\alpha}^{\beta} \delta_{\dot \alpha}^{\dot \beta}
\end{equation}
Let us introduce the $x$-factor as \cite{Arkani-Hamed:2017jhn}
\begin{equation}
x m |\lambda_3 \rangle = i p_1| \lambda_3 ], \qquad
-im [\lambda_3 | = x \langle \lambda_3| p_1  \,.
\end{equation}
The  polarization vectors for massive fields are defined as
\begin{equation} \label{pol-m}
\epsilon^\mu =\frac{\langle \lambda| \sigma^\mu |\lambda]}{\sqrt 2 m}\,, \qquad 
\epsilon = \sqrt 2\frac{| \lambda \rangle [\lambda |}{m}\,,
\end{equation}
and the polarization vectors for massless ones are
\begin{equation}
\epsilon_\mu^+ =\frac{1}{\sqrt 2}\frac{\langle q| \sigma_\mu |\lambda]}{ \langle q \lambda \rangle}\,, \qquad 
\epsilon^+ = \sqrt 2\frac{| q \rangle [\lambda |}{ \langle q \lambda \rangle}
\end{equation}
\begin{equation}
\epsilon_\mu^- =\frac{1}{\sqrt 2}\frac{\langle \lambda|  \sigma_\mu |q]}{ [\lambda q] }\,, \qquad 
\epsilon^- = \sqrt 2\frac{| \lambda \rangle [q |}{[\lambda q]}
\end{equation}
where $|q \rangle$ and $|q ]$ are arbitrary reference spinors. Their presence is a consequence of the gauge invariance. Finally, one can check the following useful identities
\begin{equation}
\frac{\sqrt 2i}{m} \epsilon_3^+ \cdot p_1 =
x\,, \quad
\frac{\sqrt 2i}{m} \epsilon_3^- \cdot p_1 =
\frac{1}{x}\,,
\end{equation}
\begin{equation}
[13] = ix \langle 1 3 \rangle\,, \quad 
[23] = -ix \langle 2 3 \rangle\,, \quad
x \langle 2 3 \rangle \langle 3 1 \rangle=
im (\langle 2 1  \rangle- [ 2 1 ])\,,
\end{equation}
\begin{equation}
p_{1, \dot \alpha}{}^{\alpha} \, p_{2, \alpha \dot \beta}
= m^2 \varepsilon_{\dot \alpha \dot \beta}
+ \frac{im}{x} \bar \lambda_{3, \dot \alpha}
\bar \lambda_{3, \dot \beta}\,,
\end{equation}
\begin{equation}
[2 |p_1|3 \rangle = -m \langle 3 2 \rangle\,,
\quad [1 |p_2| 3 \rangle = -m \langle 3 1 \rangle\,.
\end{equation}

\paragraph{Massive reference spinors.} Massive polarization tensors are gauge-invariant, but in view of the massive higher-spin gauge symmetry it might be useful to define ``gauge-variant'' polarization tensors and massive reference spinors. Let us consider as an example the gauge invariant formulation for a massive spin-$1$ field. As we discussed in  Section \ref{sec:freeBRST},
one can write the gauge invariant massive vector field
$\tilde \phi_{\alpha \dot \alpha}$ as a sum
\begin{equation}
\tilde \phi_{\alpha \dot \alpha } =
\phi_{\alpha \dot \alpha } - \frac{p_{\alpha \dot \alpha}}{m} \phi\,.
\end{equation}
The polarization vector for $\tilde \phi_{\alpha \dot \alpha }$
is given by \p{pol-m}.
The polarization vector for $\phi_{\alpha \dot \alpha }$
is obtained by
\begin{equation}
\tilde \epsilon_{\alpha \dot \alpha }= \epsilon_{\alpha \dot \alpha }
- \frac{1}{2m} p_{\alpha \dot \alpha } \phi
\end{equation}
with
\begin{equation}
\phi = \frac{1}{m} (q^{a, \alpha} \lambda_{a, \alpha}
+ \bar q^{a}_{ \dot \alpha} \bar \lambda_{a}^{\dot \alpha}
)
\end{equation}
Since the variation of the reference spinor has the form\footnote{See \cite{Witten:2003nn}
for the analogous discussion for massless vector fields.}
$\delta q^{a, \alpha} = \tau^{ ab} \lambda_{b}{}^{\alpha} $,
we see that $\delta \tilde \epsilon_{\alpha \dot \alpha } =
p_{\alpha \dot \alpha }(\tau^a_a + \bar \tau^a_a)$.
Notice, that when computing the amplitudes, one can fix the reference
spinor to an arbitrary value, for each gauge invariant field separately.
Choosing it equal to $\bar \lambda_{\dot \alpha}$,  one can make the field $\phi$ equal to zero.
Therefore, when computing  on-shell amplitudes by using the vertices
\p{sbv-1}-- \p{sbv-5} one can take into account only the $\alpha^{(i),+}_\mu$
dependence for the external fields.

\footnotesize
\providecommand{\href}[2]{#2}\begingroup\raggedright\endgroup

\end{document}